\documentstyle[11pt]{article}

\makeatletter
\@addtoreset{equation}{section}
\makeatother

\let\la=\label  
  
 \def\bd{\begin{document}} \def\ed{\end{document}}
\def\ds{\documentstyle} \let\fr=\frac \let\bl=\bigl \let\br=\bigr
\let\Br=\Bigr \let\Bl=\Bigl
\let\bm=\bibitem
\let\na=\nabla
\let\pa=\partial \let\ov=\overline
\newcommand{\beq}{\begin{equation}}
\newcommand{\eeq}{\end{equation}}
\def\ba{\begin{array}}
\def\ea{\end{array}}
\def\ft#1#2{{\textstyle{{\scriptstyle #1}\over {\scriptstyle #2}}}}
\def\fft#1#2{{#1 \over #2}}
\def\del{\partial}
\def\sst#1{{\scriptscriptstyle #1}}
\def\oneone{\rlap 1\mkern4mu{\rm l}}
\newcommand{\ho}[1]{$\, ^{#1}$}
\newcommand{\hoch}[1]{$\, ^{#1}$}
\newcommand{\bea}{\begin{eqnarray}}
\newcommand{\eea}{\end{eqnarray}}
\newcommand{\ra}{\rightarrow}
\newcommand{\lra}{\longrightarrow}
\newcommand{\Lra}{\Leftrightarrow}
\newcommand{\ap}{\alpha^\prime}
\newcommand{\bp}{\tilde \beta^\prime}
\newcommand{\tr}{{\rm tr} }
\newcommand{\Tr}{{\rm Tr} }
\newcommand{\NP}{Nucl. Phys. }
\newcommand{\tamphys}{\it Center for Theoretical Physics\\
Texas A\&M University, College Station, Texas 77843}

\newcommand{\auth}{M. J. Duff\hoch{\dagger}, H.
L\"u\hoch{\ddagger}, C. N. Pope\hoch{\ddagger}
and E. Sezgin\hoch{\dagger}}

\begin{document}

\hfill{CTP-TAMU-35/95}

\hfill{hep-th/9511162}

\vspace{20pt}

\begin{center}
{ \large {\bf SUPERMEMBRANES WITH FEWER SUPERSYMMETRIES}}
\vspace{30pt}

\auth

\vspace{15pt}

{\tamphys}

\vspace{40pt}

\underline{ABSTRACT}
\end{center}

The usual supermembrane solution of $D=11$ supergravity interpolates between
$R^{11}$ and $AdS_4 \times round~S^7$, has symmetry $P_3 \times SO(8)$ and
preserves $1/2$ of the spacetime supersymmetries for either orientation of
the round $S^7$.  Here we show that more general supermembrane solutions
may be obtained by replacing the round $S^7$ by any seven-dimensional
Einstein space $M^7$.  These have symmetry $P_3 \times G$, where $G$ is the
isometry group of $M^7$. For example, $G=SO(5) \times SO(3)$ for the
squashed $S^7$. For one orientation of $M^7$, they preserve $N/16$ spacetime
supersymmetries where $1\leq N \leq 8$ is the number of Killing spinors on
$M^7$; for the opposite orientation they preserve no supersymmetries since
then $M^7$ has no Killing spinors.  For example $N=1$ for the left-squashed
$S^7$ owing to its $G_2$ Weyl holonomy, whereas $N=0$ for the right-squashed
$S^7$. All these solutions saturate the same Bogomol'nyi bound between the
mass and charge. Similar replacements of $S^{D-p-2}$ by Einstein spaces
$M^{D-p-2}$ yield new super $p$-brane solutions in other spacetime
dimensions $D\leq 11$.  In particular, simultaneous dimensional reduction of
the above $D=11$ supermembranes on $S^1$ leads to a new class of $D=10$
elementary string solutions which also have fewer supersymmetries.

{\vfill\leftline{}\vfill
\vskip	10pt
\footnoterule
{\footnotesize
	\hoch{\dagger}	Research supported in part by NSF Grant	PHY-9411543
\vskip	-12pt} \vskip	10pt {\footnotesize
	\hoch{\ddagger}	Research supported in part by DOE
Grant DE-FG05-91-ER40633 \vskip	-12pt}}

\pagebreak
\setcounter{page}{1}

\section{Introduction}
\la{introduction}

Recent developments in non-perturbative string theory have emphasized the
need to incorporate supersymmetric extended objects of more than two
worldvolume dimensions \cite{Khuristring}: the super $p$-branes. They appear
as solitons of the fundamental string theory and may even need to be treated
as fundamental in their own right.  At the same time, there has been a
revival in the fortunes of $D=11$ dimensions: $D=11$ supergravity
\cite{Scherk,Pope} emerges as the strong coupling limit of the Type $IIA$
string \cite{Wittenvarious} and there are hints that the $D=11$
supermembrane \cite{Bergshoeff2,Stelle} on $S^1$ and the $D=10$ Type $IIA$
string theories may actually be equivalent when all the solitonic states are
taken into account \cite{Townsendeleven}. Certainly, the Green-Schwarz
action of the latter follows by simultaneous worldvolume/spacetime $S^1$
dimensional reduction of the former \cite{Howe}.  Moreover, there is a web
of interconnections between the $D=11$ supermembrane and the other four
superstring theories \cite{Minasian2,Becker,Schwarz,Horava} and it has even
been suggested that the $D=11$ supermembrane might be the fundamental theory
underlying all the others.  This therefore seems an
appropriate time to reevaluate the status of the $D=11$ supermembrane.  This
object was first described by a manifestly spacetime supersymmetric
Green-Schwarz action \cite{Bergshoeff2} whose kappa-symmetry forces the
supermultiplet of background fields $(g_{MN}, \Psi_M, A_{MNP})$ to obey the
field equations of $D=11$ supergravity. Only later was the supermembrane
recognized as a solution of the $D=11$ supergravity field equations
\cite{Stelle}, whose zero modes correspond to the physical degrees of
freedom of the Green-Schwarz action.  This contrasts with some of the other
super $p$-branes whose existence was first established by solving the
supergravity field equations and for which, in some cases, the Green-Schwarz
action is still unknown.

The usual supermembrane solution of $D=11$ supergravity interpolates between
$R^{11}$ and $AdS_4 \times round~S^7$ \cite{Gibbons}, has symmetry $P_3
\times SO(8)$, preserves $1/2$ of the spacetime supersymmetries  and admits
$8$ bosonic plus $8$ fermionic zero modes \cite{Stelle} (for either
orientation of the round $S^7$.)  Here we show that more general
supermembrane solutions may be obtained by replacing the round $S^7$ by any
seven-dimensional Einstein space $M^7$.  These have symmetry $P_3 \times G$,
where $G$ is the isometry group of $M^7$. For example, $G=SO(5) \times
SO(3)$ for the squashed $S^7$.  For one orientation of $M^7$, they preserve
$N/16$ spacetime supersymmetries where $1\leq N \leq 8$ is the number of
Killing spinors on $M^7$; for the opposite orientation they preserve no
supersymmetries since then $M^7$ has no Killing spinors.  For example $N=1$
for the left-squashed $S^7$ owing to the $G_2$ Weyl holonomy, whereas $N=0$
for the right-squashed $S^7$ \cite{awada}. All these solutions saturate the
same Bogomol'nyi bound between the mass and charge.  Similar replacements of
$S^{D-p-2}$ by Einstein spaces $M^{D-p-2}$ yield new super $p$-brane
solutions in other spacetime dimensions $D\leq 11$.

We begin in section \ref{solution} by showing in general how new $p$-brane
solutions of supergravity in various spacetime dimensions $D$ may be
obtained by replacing the unit $S^{D-p-2}$ which appears in the transverse
space line element by general Einstein spaces $M^{D-p-2}$ with the same
cosmological constant.  As an important special case we find new elementary
membrane ($p=2$) solutions of $D=11$ supergravity. In section \ref{bound}
we derive a Bogomol'nyi bound between the mass per unit area of the membrane
and the central charges appearing in the $D=11$ supersymmetry algebra. All
the elementary membrane solutions, both old and new, saturate the bound. The
zero modes are analyzed in section \ref{modes}.  In contrast to the
round $S^7$, however, these new membrane solutions exhibit no translation
zero modes or their fermionic partners. The non-supersymmetric orientation
yields just $N$ fermionic zero modes while the supersymmetric orientation
yields no zero modes at all. The new elementary membrane solutions exhibit
$\delta$-function singularities at their location and hence require the
introduction of membrane sources.  As discussed in section \ref{sources},
the same ansatz that worked for the usual membrane continues to work for the
new membranes. Another interesting special case provides the focus of
section \ref{strings}, where simultaneous dimensional reduction of the
above $D=11$ supermembranes on $S^1$ leads to a new class of $D=10$
elementary string solutions which also have fewer supersymmetries. Having
replaced the round $S^{D-p-2}$ by Einstein spaces $M^{D-p-2}$ with the
same positive cosmological constant, the question arises as to whether we
can find new super $p$-brane solutions when $M^{D-p-2}$ is Ricci flat. As
discussed in section \ref{othersol}, the answer is yes, and we give examples
such as the seven-dimensional Joyce manifolds which also have $G_2$ holonomy
and hence yield $N=1$.  In this case, one finds $N$ bosonic and $N$
fermionic zero-modes.  These Ricci-flat manifolds do give rise to $p$-branes
with rather unusual asymptotic behavior, however.  We finish with some
concluding remarks in section \ref{conclusion}.

\section{New $p$-brane solutions}
\la{solution}

      Our principle focus in this paper is to find new supermembrane
solutions of $D=11$ supergravity. However, we shall begin by considering the
more general case of a $p$-brane in $D$ dimensions.  As in the previously
discussed elementary and solitonic $p$-brane solutions \cite{Lublack,lpss},
the relevant fields that participate in the solutions are the metric, a
dilaton and an $n$'th rank antisymmetric tensor.  The relevant part of the
bosonic Lagrangian is given by
\beq
{\cal L} = \frac{1}{2\kappa^2}(e R - \ft12 e(\del\phi)^2 - {1\over 2\, n!} e
e^{-a \phi} F_n^2)\ ,
\label{boslag}
\eeq
where $e=\sqrt{-g}$ is the determinant of the vielbein. The constant $a$
takes different values for supergravities in diverse dimensions and for
different values of $n$.  It is conveniently parametrized as
\beq
a^2=\Delta - {2(n-1)(D-n-1)\over D-2} \ .
\eeq
If $F_n$ represents the field strength of the antisymmetric tensor that
couples to the $p$-brane worldvolume via the Wess-Zumino term, then always
$\Delta=4$ \cite{Lublack}.  However, for the purpose of finding $p$-brane
solutions, $F_n$ might sometimes represent some different linear combination
of the original $n$'th rank antisymmetric tensors in the supergravity
theory. Under these circumstances, one can have $\Delta< 4$
\cite{lpss,Rahmfeld}.

We shall take the ansatz for the canonical metric for the $D$ dimensional
spacetime to be
\beq
ds^2 = e^{2A} dx^\mu dx^\nu \eta_{\mu\nu} +
       e^{2B} \Big( dr^2 + r^2 d\bar s^2\Big)\ ,\label{metricform}
\eeq
where $x^\mu$ ($\mu = 0, \ldots, d-1$) are the coordinates of the
$(d-1)$-brane world volume.   The remaining coordinates of the $D$
dimensional spacetime are $r$ and the coordinates on a $(D-d-1)$ dimensional
space $M^{D-d-1}$ whose metric is $d\bar s^2$.  This metric is taken to be
an Einstein metric with $\bar R_{\alpha\beta} = \Lambda
\delta_{\alpha\beta}$ (here $\alpha,\beta,\cdots$ denote tangent-space
indices on $M^{D-d-1}$.) The functions $A$ and $B$ depend only on $r$.
In the elementary and solitonic solutions that have been discussed
previously, $M^{D-d-1}$ has been taken to be the usual unit $(D-d-1)$
sphere.  Here we investigate the properties of $p$-brane solutions with
other choices of $M^{D-d-1}$.

     A convenient choice of vielbein for the metric (\ref{metricform}) is
$e^{\underline\mu} = e^{A} dx^\mu$, $e^r = e^B dr$ and $e^\alpha = r e^B
\bar e^\alpha$, where $\bar e^\alpha$ is the vielbein for the Einstein
metric $d\bar s^2$.   The corresponding spin connection is
\beq
\omega^{r\underline \mu} = - e^{-B} A' e^{\underline \mu}\ ,\quad
\omega^{r\alpha} = -e ^{-B}(B' +\fft1{r}) e^\alpha\ ,\quad
\omega^{\alpha\beta} = \bar \omega^{\alpha\beta}\ ,\label{spinconn}
\eeq
where $\bar \omega^{\alpha\beta}$ is the spin connection for the vielbein
$\bar e^\alpha$, and $A'={\del A\over \del r}$, {\it etc.}    The Ricci
tensor can be easily obtained; its tangent-space components are
\bea
R_{\underline\mu\underline \nu} &=& - e^{-2B} \Big( \tilde d A'B' + A'' +
d A'^2 + {\tilde d + 1\over r} A'\Big )\eta_{\underline\mu\underline\nu}\ ,
\nonumber\\
R_{rr} &=& -e^{-2B} \Big ( (\tilde d + 1) B'' - d A' B' + d A'' - dA'^2
+ {\tilde d + 1\over r} B' \Big)\ ,\label{ricci}\\
R_{\alpha\beta} &=& -e^{-2B} \Big( B'' + \tilde d B'^2 + d A' B'  +
{2\tilde d +1\over r} B' + {d\over r} A' + {\tilde d-\Lambda\over r^2}\Big)
\delta_{\alpha\beta}\ .\nonumber
\eea
There are two discrete possibilities for the metric on the Einstein space
$M^{D-d-1}$, namely $\Lambda$ zero or non-zero.  The case when $\Lambda =0$
shall be discussed in section (\ref{othersol}).  In this and the next two
sections we shall discuss only the case with $\Lambda$ positive.  Without
loss of generality, we can choose the same cosmological constant $\Lambda =
\tilde d$ as that of the unit $(\tilde d+1)$ sphere, the Ricci tensor
(\ref{ricci}) for the metric $ds^2$ is identical to that for the case when
$M$ is the unit sphere. Thus the bosonic equations giving rise to the
$p$-brane solutions will be identical to those for the case when $M$ is the
unit sphere.

For the elementary $p$-brane solutions, the ansatz for the antisymmetric
tensor is given in terms of its potential, and it takes the form
\beq
A_{\mu_1\ldots\mu_{n-1}} = \epsilon_{\mu_1\ldots\mu_{n-1}} e^C
\ ,\label{eleans}
\eeq
and hence
\beq
F_{r\mu_1\ldots\mu_{n-1}} = \epsilon_{\mu_1\ldots\mu_{n-1}} (e^C)' \ ,
\label{eleans2}
\eeq
where $C$ is a function of $r$ only, and the indices on $A$ and $F$ are
world indices.  Here and throughout this paper $\epsilon_{\sst{M\cdots N}}$
and $\epsilon^{\sst{M\cdots N}}$ have purely numerical components $\pm 1$ or
$0$.  The dimension of the world volume is given by $d=n-1=p+1$ for the
elementary $p$-brane solutions.

     The elementary $p$-brane solution to the equations of motion following
from (\ref{boslag}) is given by
\bea
e^A &=& \Big(1+{k\over r^{\tilde d}}\Big)^{-2{\tilde d}\over {\Delta(D-2)}}
\ , \qquad\qquad e^C=\frac{2}{\sqrt \Delta }\Big(1+{k\over
r^{\tilde d}}\Big)^{-1}\ , \nonumber \\
B &=& -{d\over {\tilde d}}\, A\ , \qquad\qquad\qquad\qquad
\phi={a(D-2)\over {\tilde d}}A \ ,  \label{gsol}
\eea
where $k$ is an arbitrary integration constant and we have set the asymptotic
value of the dilaton equal to zero.  Note that the $p$-brane solution
(\ref{gsol}) is  consistent with the case where the dilaton is absent in the
supergravity  theory, {\it i.e.\ }$a=0=\phi$.  Note also that when
$M^{D-d-1}$ is the usual unit sphere, the metric (\ref{metricform}) is
asymptotic to $D$-dimensional Minkowski spacetime as $r$ approaches
infinity. On the other hand, if $M^{D-d-1}$ is any other Einstein space,
then the metric (\ref{metricform}) is no longer asymptotic to Minkowski
spacetime, although it is still asymptotically flat. In fact, the metric on
the (D-d)-dimensional transverse space is asymptotic to  a Ricci-flat metric
on a generalized cone:
\beq
ds_{D-d}^2 = dr^2 + r^2d\bar s^2\ .\label{cone}
\eeq
This metric has a conical singularity at the apex $r=0$, except when $d\bar
s^2 $ is the usual round sphere metric.  In \cite{conehead}, the metric
(\ref{cone}) was rounded off at its apex to a `bolt of the second kind',
giving a complete Ricci-flat metric.  In our paper,  the singularity of
$r=0$ is eliminated.  One can see this by first going to the $\sigma$-model
metric \cite{Lublack}
\beq
ds^2(\sigma-model)=e^{- a\phi/{\tilde d}}ds^2(canonical)
\eeq
where the canonical metric is given in (\ref{metricform}), and then noting
that the $\sigma$-model metric there approaches $(AdS)_{d+1}\times M^{\tilde
d +1}$ for $\tilde d \neq 2$ or $(Mink)_{d+1} \times M^3 $ for $\tilde d =2$
\cite{Gibbons}.  Note that the translational invariance of the
$8$-dimensional metric (\ref{cone}) is absent except in the case of the
round $S^7$.  This will prove important when we come to analyse the
zero-modes in section (\ref{modes}).  Of course, the complete metric still
has the Poincar\'e symmetry in the world volume, since the only modification
is in the structure of the transverse space.

  A particular case which is important to us in the present paper is the
elementary membrane solution of $D=11$ supergravity. In this case, there is
no dilaton, and a single fourth-rank antisymmetric tensor. In fact, it fits
into the general pattern of solutions given above, with $d=3$, $\tilde d=6$,
$\Delta=4$ and hence $a=0$. From (\ref{gsol}) we therefore have
\beq
e^A = \Big( 1+ {k\over r^6}\Big)^{1/3}\ , \qquad
e^C=\Big( 1+{k\over r^6}\Big)^{-1}\, \qquad
B=-\ft12 A\ ,\label{d11sol}
\eeq
Note that the replacement of the round $S^7$ by the Einstein space $M^7$ has
not changed the singularity structure discussed in \cite{Gibbons}: by
introducing the Schwarzschild-like coordinate $\rho$ given by
\beq
r^6=\rho^6-k^6
\eeq
one may see that there is an horizon at $\rho=k$ and that the metric may be
analytically continued through the horizon up to a curvature singularity at
$\rho=0$.

\section{Supersymmetry and Bogomol'nyi bound}
\la{bound}

      In the previous section, we obtained new $p$-brane solutions by
replacing the usual round sphere that foliate the transverse space with a
more general Einstein space.   In particular, we obtained new elementary
membrane solutions in $D=11$ supergravity.   In this section we shall
investigate the supersymmetry of these solutions.

     The gravitino transformation rule in $D=11$ supergravity with bosonic
background is given by
\beq
\delta \psi_{\sst A} = D_{\sst A} \epsilon
 -\ft1{288} (\Gamma_{\sst A}{}^{\sst{B C DE}} - 8 \delta^{\sst B}_{\sst A}
\, \Gamma^{\sst{CDE}}) F_{\sst{BCDE}} \epsilon \equiv \hat D_{\sst A}
\epsilon\ , \label{susy1}
\eeq
where $\sst A, \sst B, \ldots$ are tangent space indices. Decomposing these
indices in terms of $\sst A=(\underline \mu, r, \alpha)$, we first make a
$3+1+7$ split of gamma matrices:
\beq
\Gamma^{\underline{\mu}} = \gamma^{\underline{\mu}} \otimes \sigma_3 \otimes
\oneone\ ,\quad \Gamma^r=\oneone \otimes \sigma_1 \otimes \oneone\ ,
\quad \Gamma^\alpha = \oneone\otimes \sigma_2 \otimes \gamma^\alpha\ .
\label{gammadecom}
\eeq
Substituting the solutions from the previous section, the gravitino
transformation rule (\ref{susy1}) becomes
\bea
\delta \psi_{\underline\mu} &=&
e^{-A} \del_\mu \epsilon -
\ft12 e^{-B} A' \gamma_{\underline\mu} \otimes \sigma_1
(\sigma_3 -\oneone) \otimes \oneone \epsilon\ ,\nonumber\\
\delta \psi_r &=& e^{-B} \epsilon' -\ft12 e^{-B} A'\, \oneone \otimes
\sigma_3 \otimes \oneone\, \epsilon\ ,\label{susy2}\\
\delta \psi_\alpha &=& \fft{1}{r} e^{-B} \Big (\oneone \otimes \oneone
\otimes {\bar D}_\alpha \epsilon -\fft{\rm i}2 \oneone \otimes \sigma_3 \otimes
\gamma_\alpha \epsilon\Big) +\fft{\rm i}4 e^{-B} A'\, \oneone\otimes
(\sigma_3 -\oneone) \otimes \gamma_\alpha \epsilon\ .\nonumber
\eea
It is easy to see that these variations all vanish provided that
\beq
\epsilon = e^{\fft12 A}\,\epsilon^0_{+} \otimes\eta_{-}\ , \qquad
\oneone\otimes \sigma_3 \epsilon^0_{+}= \epsilon^0_{+}\ ,
\label{spinsol}
\eeq
where $\eta_{-}$ is a Killing spinor on the Einstein space $M^7$, satisfying
\beq
\bar D_\alpha \eta_{-} - \fft{\rm i}2 \gamma_\alpha \eta_{-} = 0\ .
\label{killing}
\eeq
Thus if there are $N$ Killing spinors $\eta_{-}$ satisfying (\ref{killing})
on the Einstein space $M^7$, then $N/16$ of the $D=11$ spacetime
supersymmetry is preserved. For example, if $M^7$ is the usual round seven
sphere, for which there are eight Killing spinors satisfying
(\ref{killing}), we recover the usual half breaking of supersymmetry, which
was found in \cite{Stelle}.   Other examples are provided by the
squashed seven sphere, which has $N=1$ \cite{awada}; the $M^{pqr}$ spaces,
which have $N=2$ \cite{fre}; and the $N^{pqr}$ spaces, which have $N=1$ or
$N=3$ \cite{npqr}.  These examples give rise to new supermembrane solutions
in $D=11$ supergravity which preserve $1/16$, $1/8$ and $3/16$ of the
original $D=11$ supersymmetry.  If, on the other hand we consider the same
spaces with their orientations reversed then, with the exception of the
round $S^7$, all supersymmetries are broken.  This is the {\it skew-whiffing
theorem} \cite{Pope}; the round $S^7$ admits eight solutions of $\bar
D_\alpha \eta_{-} -\ft12{\rm i} \gamma_\alpha \eta_{-} =0$ and eight
solutions of $\bar D_\alpha \eta_{+} + \ft12 {\rm i} \gamma_\alpha \eta_{+}
= 0$, but any other non-Ricci-flat $M^7$ can admit solutions of either the
former or the latter equation (depending on orientation) but not both. It is
worth remarking that except for the round $S^7$, when the membrane is
absent, {\it i.e.}\ $k=0$, the metric (\ref{metricform}) and (\ref{gsol}) is
not Minkowskian, and already breaks $(1- N/16)$ of the $D=11$ supersymmetry.
The remaining supersymmetry is either broken or unbroken by the presence of
the membrane, depending on the orientation of $M^7$.  We shall see in
section 4 that this leads to the the unusual counting of zero modes of the
solution.

     It is instructive to examine the relation between the mass per unit
area, $m$, and the Page charge \cite{page}, $Z$, which gives rise to a
central charge in the supersymmetry algebra:
\beq
\{Q, Q\} = \Gamma^{\sst A} P_{\sst A} + \Gamma^{\sst {AB}} U_{\sst{AB}} +
\Gamma^{\sst{ABCDE}} V_{\sst{ABCDE}} \ .\label{qqalg}
\eeq
The elementary solutions will contribute to the $U$-type central charge
while the solitonic solutions, which we have not considered so far, will
contribute to the $V$-type central charge.  To calculate the central
charges, we need to compute the supercharges and then their anticommutator.

    The supersymmetric variation of $D=11$ supergravity takes the form
$\delta I = \int \hat D_{\sst M} \bar \epsilon J^{\sst M}$, where the
Noether current is given by $J^{\sst M} = \Gamma^{\sst{MNP}} \hat D_{\sst N}
\psi_{\sst P}$, modulo the bosonic field equations.  Thus the supercharges
per unit membrane area are given by
\beq
Q_{\epsilon} = \int_{\del\Sigma} \bar\epsilon
\Gamma^{\sst{ABC}} \psi_{\sst C} d\Sigma_{\sst{AB}}\ ,\label{supercharge}
\eeq
where $\Sigma$ is an eight dimensional space-like surface.  When
$r\longrightarrow \infty$, we take $\epsilon$ to be such that $\epsilon
\longrightarrow \epsilon_0\otimes \eta$, where $\epsilon_0$ is a constant
four-component spinor on the space of $(x^{\mu}, r)$ and $\eta$ is a Killing
spinor on the Einstein space $M^7$, satisfying (\ref{killing}). The
commutator of the resulting conserved supercharges is given by
\beq
[Q_{\epsilon_1}, Q_{\epsilon_2}]  =\delta_{\epsilon_1}  Q_{\epsilon_2}
=\int_{\del\Sigma} N^{\sst{AB}} d\Sigma_{\sst{AB}}\ ,\label{commut}
\eeq
where $N^{\sst{AB}} = \bar\epsilon_1 \Gamma^{\sst{ABC}} \delta
\psi_{\sst C}$.  Using (\ref{susy1}), we obtain
\beq
N^{\sst{AB}} = \bar \epsilon_1 \Gamma^{\sst{ABC}} D_{\sst C} \epsilon_2
+ \ft18 \bar\epsilon_1 \Gamma^{\sst{C_1
C_2}} \epsilon_2 F^{\sst{AB}}{}_{\sst{C_1C_2}} +
\ft1{96} \bar \epsilon_1 \Gamma^{\sst{ABC_1\ldots C_4}} \epsilon_2
F_{\sst{C_1\ldots C_4}}\ .\label{nest}
\eeq
Substituting elementary membrane solution (\ref{d11sol}) and gamma matrix
decomposition (\ref{gammadecom}), the last term of (\ref{nest}) vanishes in
(\ref{commut}) since only the $d \Sigma_{0r}$ component of the area element
contributes.   Thus we obtain the Bogomol'nyi matrix ${\cal M}$:
\beq
\int_{M\, {\rm at}\, r\rightarrow \infty} N^{0r} r^7 d\Omega_7 =-2/3
\epsilon_1^{\dagger} {\cal M} \epsilon_2\, \omega_7\ ,
\eeq
where $\omega_7$ is the volume of $M$, and
\beq
{\cal M} = m \oneone + Z \sigma_3 \ ,\label{calm}
\eeq
where we have suppressed the common-factor gamma matrices in the world
volume, $m$ is the mass per unit membrane area corresponding to the first
term in (\ref{nest}) and $Z$ is the central charge corresponding to the
second term in (\ref{nest}).    In fact, $Z$ is precisely equal to the
Page charge defined by $\int *F/\omega_7$, with $F = 6k/r^7 dr\wedge
dx^0\wedge dx^1\wedge dx^2$ asymptotically at large $r$.  With the new
solutions obtained in this paper, it is easy to see that $m=k= Z$. The
matrix ${\cal M}$ has two eigenvalues $\{m -Z, m+Z\}=\{0, 2k\}$.  The zero
eigenvalue implies that $1/16$ of the $D=11$ supersymmetry is preserved for
each Killing spinor.  Thus if there are $N$ Killing spinors satisfying
(\ref{killing}) on the Einstein space $M^7$, then $N/16$ of the spacetime
supersymmetry is preserved, which is the same conclusion we obtained at the
beginning of this section by looking explicitly at the supersymmetry
transformation rules.

     For an arbitrary configuration, it can be shown that there is a
Bogomol'nyi bound $m\ge |Z|$.  See for example \cite{Dabholkar}.  Thus we
see that this bound is saturated for any of the elementary supermembrane
solutions obtained in this paper.  It is interesting to note that the
mass-charge ratio is independent of the choice of the Einstein space $M^7$.
Since the choice of the Einstein space determines how much of the $D=11$
dimensional supersymmetry is broken, it follows that for these types of
solutions, the mass-charge ratio is independent of how much of the
supersymmetry is broken. In this respect, the solutions presented in this
paper are qualitatively different from previous examples of p-brane
solutions breaking more than half the supersymmetries, such as the extreme
black fourbrane and sixbrane \cite{Guven} of the $N=1,D=11$ theory  (which
break $3/4$ and $7/8$, respectively), the octonionic string soliton
\cite{Harveystrominger} of the $N=1,D=10$ theory (which breaks $15/16$), the
double instanton string soliton  \cite{Khuri2} of the $N=1,D=10$ theory
(which breaks $3/4$), the elementary particle solution \cite{lps} of the
$N=2,D=9$ theory which breaks $3/4$, the dyonic string \cite{Rahmfeld2} of
the $N=2,D=6$ theory which breaks $3/4$, the extreme $a=0$ and
$a=1/\sqrt{3}$ black holes of the $N=4,D=4$ theory which both break $3/4$
\cite{Kalloshpeet,Rahmfeld1}, and the string solitons of the $N=4,D=4$
theory which break $3/4$ and $7/8$ \cite{Rahmfeld2}. In the case of the
double instanton string and the octonionic string, for example, the mass per
unit length actually diverges, in contrast to the elementary strings
discussed in section (\ref{strings}).  In supergravities with $N$-extended
supersymmetry, there will be  several central charges $Z_1$,
$Z_2$...$Z_{N/2}$ and the mass is bounded by $Z_{max}$.  The number of
supersymmetries preserved by the solution will then be related to the number
of $Z$s that are equal to $M$. In these examples, the Bogomol'nyi bounds are
different from the solutions breaking just $1/2$ of the supersymmetries
because one changes the value of $\Delta$. In this paper, we obtain fewer
supersymmetries by keeping $\Delta$ fixed but replacing $S^{D-p-2}$ by
$M^{D-p-2}$.  In fact as we saw earlier, $M^{D-p-2}$ has already partially
broken the $D=11$ supersymmetry.  The presence of the $p$-brane itself
breaks none or all of the remaing supersymmetry depending on the
orientation.

The new super $p$-brane solutions discussed in this section arise as a
consequence of the existence of Einstein spaces $M^{D-p-2}$ other than round
spheres that admit Killing spinors.   Such spaces seem to be known only for
the dimensions $\ge 5$.   Thus in $D$ dimensions, there could only be new
solutions with world volume dimension  $\le D-6$.

\section{Zero Modes}
\la{modes}

Each broken supersymmetry transformation of a $p$-brane solution gives rise
to a corresponding fermionic Goldstone zero-mode. There will also be bosonic
zero modes associated with the breaking of local bosonic gauge symmetries.
If supersymmetry remains partially unbroken by the solution, then the
fermionic and bosonic  zero modes will form supermultiplets of the unbroken
symmetry. These  multiplet furnish a linear realization of the unbroken
supersymmetry. The  full supersymmetry of the $D$ dimensional theory is
still realised of course, albeit nonlinearly.

In the case of the round $S^{D-p-2}$, these zero-modes will include the
$(D-d)$ translational zero modes corresponding to the breaking of the
$D$-dimensional translational symmetry down to the $d$-dimensional world
volume translations. However, our new solutions are dramatically different
in this respect: the spacetime even in the absence of the $p$-brane (i.e.
when $k=0$) has translational invariance only in the worldvolume directions
and not in the transverse directions.  Consequently there are no Goldstone
zero-modes corresponding to the breaking of the $(D-d)$-dimensional
translations. In fact, as we shall see, in the case where there is some
residual supersymmetry, there are no zero-modes at all! This presumably
means that the corresponding Green-Schwarz action is that of a $p$-brane in
$p+1$ dimensions which has no continuous physical degrees of freedom.

     We shall first consider the fermionic zero modes.  Note that unlike the
case of the round $S^7$, even when the membrane solutions are absent, {\it
i.e.} $k=0$, there are only $N<8$ components of supersymmetry, where $N$ is
the number of Killing spinors of $M^7$.   The gravitino supersymmetry
transformation in the backgrounds of the new elementary supermembrane
solutions are given by (\ref{susy2}).  As we shall see now, the
supersymmetry of the solution depends on the orientation of $M^7$, which we
denote by $M_{\pm}^7$. For $M_{-}^7$, where the Killing spinors satisfy
$\bar D_\alpha \eta_{-} -\ft12 {\rm i} \gamma_\alpha \eta_{-}=0$, the
variations (\ref{susy2}) vanish for spinors $\epsilon$ satisfying
(\ref{spinsol}) and (\ref{killing}), which includes a chirality condition
and Killing-spinor condition.  They correspond to the $N$ unbroken
supersymmetry generators.   The fermionic zero modes would correspond to the
spinors $\epsilon$ for which the variations (\ref{susy2}) do not vanish.
However, these ``zero modes'' are not normalizable since the first term of
$\delta \psi_\alpha $ in (\ref{susy2}) does not vanish fast enough as $r
\rightarrow \infty$.  Thus in this case,  there are no normalizable
fermionic zero modes.  In fact this is understandable since the membrane
solution preserves all $N$ components of supersymmetry.   It follows by
supersymmetry that there are no bosonic zero modes either in this case.

      For the case of $M_{+}^7$, for which the Killing spinors satisfy
$\bar D_\alpha \eta_{+} + \ft12 {\rm i} \gamma_\alpha \eta_{+}=0$,  there is
no spinor $\epsilon$ such that the gravitino supersymmetry variations
(\ref{susy2}) vanish.  Thus in this case, the membrane solution breaks all
$N$ components of supersymmetry which is present when the membrane is
absent. Accordingly there are $N$ normalizable fermionic zero modes, given
by $\epsilon = e^{-\ft12 A} \epsilon^0_{-} \otimes \eta_{+}$ and $\oneone
\otimes \sigma_3 \epsilon^0_{-} = -\epsilon_{+}^0$, where $\eta_{+}$ is any
of the $N$ Killing spinors on $M^7_{+}$.   In this non-supersymmetric
membrane solution, a separate counting of the bosonic zero modes is
necessary.  Since skew-whiffing seems not to change the bosonic symmetries,
we would expect that there are no bosonic zero modes in this case either.

\section{Including the sources}
\la{sources}

The $D=11$ supermembrane solution (\ref{d11sol}) does not solve the field
equations everywhere because of the appearance of $\delta$-function
singularities at $r=0$. Instead of
\beq
{\Box}_8e^{-C}=0
\eeq
for example, we have
\beq
{\Box}_8e^{-C}=-6k\Omega_7\delta^8({\underline r})
\eeq
In order that (\ref{d11sol}) be solutions
everywhere it is necessary that the pure supergravity equations be augmented by
source terms.  This source is, of course, just the supermembrane itself.  Here,
however, we encounter the problem that the covariant Green-Schwarz action for
the new membranes is unknown.  We know the correct action describing just
the usual $8$ translation zero modes and their superpartners
\cite{Bergshoeff2}. In the case of the new supermembranes with no zero-modes,
the action is presumably just a truncation of this. In the
non-supersymmetric case, there will be extra terms.
Nevertheless, the same ansatz that was used in the case of the round $S^7$
\cite{Stelle} continues to solve the field equations. It is not difficult to
verify that the correct source terms for the antisymmetric tensor and Einstein
equations, as well as the membrane equations, are given by choosing the static
gauge choice
\begin{equation} X^{\mu}=\xi^{\mu},\qquad\qquad\ \mu=0,1,2.
\end{equation}
where $X^{\mu}$ $(\mu =0,1,2)$ are the first three spacetime coordinates,
and $\xi^i$ $(i=0,1,2)$ are the worldvolume coordinates, and then setting
all the other zero mode variables to zero.  The inclusion of the source now
requires, as usual \cite{Stelle}, that the constant $k$ be given by
\begin{equation}
k=\frac{\kappa^2T}{3\Omega_7}\ ,
\end{equation}
where $T$ is the membrane tension.

\section{New elementary string solutions}
\la{strings}

An important special case of new $D<11$ solutions is provided by the elementary
strings obtained by simultaneous dimensional reduction of the new
$D=11$ supermembranes.  The logic proceeds in just the same way as for the
round $S^7$ case. Let us denote all $D=11$ variables by a carat, and then make
the ten-one split
\begin{equation}
\hat x^{\hat M}=(x^M,x^2),\qquad\qquad\ M=0,1,3,...10.
\label{reduce}
\end{equation}
Then the solutions for the $D=10$ canonical metric $g_{MN}$, $2$-form
$B_{MN}$ and dilaton $\phi$ follow from
\[
\hat g_{MN}=e^{-\phi/6}g_{MN},
\]
\[
\hat g_{22}=e^{4\phi/3},
\]
\begin{equation}
\hat A_{012}=B_{01}.
\label{string}
\end{equation}
{}From (\ref{d11sol}), we see that they are given by (\ref{gsol}) in the
special
case $d=2,D=10$ and, like the elementary solution of \cite{Dabholkar}, will
solve the heterotic and Type $IIB$ superstring equations as well as Type
$IIA$.  Moreover, like the elementary solution of \cite{Dabholkar}, they will
have finite mass per unit length, in contrast to the octonionic string
soliton \cite{Harveystrominger} and the double instanton string soliton
\cite{Khuri2}. The physical interpretation of such (non-critical?) strings
remains obscure.

\section{Further solutions}
\la{othersol}

Both the elementary and solitonic $p$-brane solutions considered above have
an interesting generalization in which the Einstein space $M^{D-p-2}$ has
zero rather than positive cosmological constant $\Lambda$.  The ansatz for
the antisymmetric tensor is the same as that for the previously discussed
solutions, given by  (\ref{eleans}) and (\ref{eleans2}).   It is convenient
to express the metric ansatz (\ref{metricform}) in terms of $ds^2 = e^{2A}
dx^\mu dx^{nu}\eta_{\mu\nu} + e^{2B} (dr^2 + d\bar s^2)$, where $d\bar s^2$
is the metric for a Ricci-flat Einstein space $M^{D-p-2}$. The spin
connection and Ricci tensors are given by  (\ref{spinconn}) and
(\ref{ricci}) respectively with all the explicit $r$ dependent terms
dropped.  The elementary $p$-brane solution to the equations of motion
following from (\ref{boslag}) is given by
\bea
e^{A} &=& (1-\lambda r)^{-1/c}\ , \qquad e^{C} = \fft{2}\Delta
\sqrt{2 c\tilde d\over D-2} \Big(1-\lambda r\Big)^{-\Delta
(D-2)/(2c\tilde d)}\ ,\nonumber\\
B&=&-\fft{d}{\tilde d} A\ , \qquad \phi = {a (D-2)\over \tilde d} A\ ,
\label{flatsol}
\eea
where $c = 3d - \Delta (D-2)/{\tilde d}$.   In particular in $D=11$, the
solutions are given by
\beq
ds^2 = (1-\lambda r)^{-2/3} dx^\mu dx^\nu\eta_{\mu\nu} +
(1-\lambda r)^{1/3} (dr^2 + d\bar s^2)\ ,\qquad
(e^C) = { 1\over 1-\lambda r}\ .\label{rflatsol}
\eeq

     The analysis of the supersymmetry of these membrane solutions in
Ricci-flat manifolds is similar to that for the supermembrane solutions
discussed in the previous sections.   The amount of supersymmetry that is
preserved depends on the number of Killing spinors (which on a Ricci-flat
space are the same as covariantly constant spinors) in the Ricci flat
manifold $M$. For example, in $D=11$, if $M$ admits $N$ such spinors, then
$N/16$ of the $D=11$ supersymmetry is preserved.   An interesting example for
$M$ is provided by Joyce manifolds, which are compact Ricci-flat seven
manifolds with $G_2$ holonomy \cite{joyce}. (In fact they are three-torus
bundles over $K3$.) Accordingly, they admit one covariantly constant spinor,
and thus $1/16$ of the original supersymmetry is preserved.   Other examples
are $T^7$, which has $N=8$; $K3\times T^3$, which has $N=4$; and $Y\times
S^1$ where $Y$ is a Ricci-flat Calabi-Yau space, which has $N=2$.

The discussion of the zero modes is different from that for the
solutions discussed in the previous sections, however, since the {\it
skew-whiffing theorem} ceases to apply in the Ricci-flat case \cite{Pope}.
Nevertheless, we shall follow the same logic that we used to discuss the zero
modes for the membrane solutions with $M^7$ non-Ricci-flat in section 4,
and compare the symmetries with and without a membrane.  For
Ricci-flat $M^7$ , the supersymmetry transformation rule for the gravitino
is given by (\ref{susy2}) except that the first term in $\delta  \psi_\alpha$
turns into $e^{-B}\,\oneone \otimes \oneone \otimes \bar  D_\alpha \epsilon$.
Thus when the membrane is absent, {\it i.e.}\ $\lambda = 0$, there are $2N$
components of the 11-dimensional supersymmetry that are preserved,
corresponding to $N$ Killing spinors $\epsilon^0_{+}\eta$ and $N$ Killing
spinors $\epsilon^0_{-}\eta$, where $\bar D_\alpha \eta = 0$. It can be seen
from (\ref{susy2}) that the variations are all proportional to
$\lambda$ except for the first term in $\delta\psi_\alpha$, which also
vanishes for the above $2N$ Killing spinors.  However, when $\lambda \ne 0$,
the variations (\ref{susy2}) vanish only for the $N$ Killing spinors
$\epsilon_{+}^0\eta$.  Thus the presence of the membrane breaks further half
of the supersymmetry, giving rise to $N$ fermionic zero modes. The
supersymmetry of the membrane then dictates that the solution must have $N$
bosonic zero modes.

\section{Conclusions}
\la{conclusion}

We have seen that the usual super $p$-brane solutions of supergravity are
merely special cases of more general solutions which can have fewer
supersymmetries.  The construction of a Green-Schwarz action for these new
$p$-branes remains an interesting unsolved problem.  Curiously, the
superfivebrane in $D=11$, which is conjectured to be dual to the
supermembrane, seems not to admit new non-Ricci-flat versions with fewer
supersymmetries since $S^4$ is the only Einstein $M^4$ known to admit
Killing spinors.  The situation is different in the case where $M^4$ is
Ricci flat, since then $K3$ provides a supersymmetric example. It also
remains to be seen how these new objects fit into the already bewildering
mesh of interconnections between $p$-branes and strings.

\section*{Acknowledgements}

We have benefited from conversations with Joachim Rahmfeld and Andy
Strominger.

\pagebreak

\end{document}